\documentclass{emulateapj}

\usepackage{amssymb}
\usepackage{amsmath}
\usepackage{hyperref}

\slugcomment{Accepted for publication in {\it{The Astrophysical Journal Letters}}}
\shortauthors{Smith et al.}
\shorttitle{The superluminous supernova \taz}

\newcommand{\nickel}{$^{56}$Ni}
\newcommand{\msolar}{\ensuremath{\mathrm{M}_{\sun}}}
\newcommand{\rsolar}{\ensuremath{\mathrm{R}_{\sun}}}
\newcommand{\mejecta}{\ensuremath{\mathrm{M}_{\mathrm{ej}}}}
\newcommand{\mnickel}{\ensuremath{\mathrm{M}_{\mathrm{Ni}}}}
\newcommand{\taz}{DES14X3taz}

\begin{document}

\title{DES14X3taz: A Type I Superluminous Supernova Showing a Luminous, Rapidly Cooling Initial Pre-Peak Bump}
\author{
M.~Smith\altaffilmark{1,$\star$},
M.~Sullivan\altaffilmark{1},
C.~B.~D'Andrea\altaffilmark{1,2},
F.~J.~Castander\altaffilmark{3},
R.~Casas\altaffilmark{3},
S.~Prajs\altaffilmark{1},
A.~Papadopoulos\altaffilmark{2,4},
R.~C.~Nichol\altaffilmark{2},
N.~V.~Karpenka\altaffilmark{1},
S.~R.~Bernard\altaffilmark{5},
P.~Brown\altaffilmark{6},
R.~Cartier\altaffilmark{1},
J.~Cooke\altaffilmark{7},
C.~Curtin\altaffilmark{7},
T.~M.~Davis\altaffilmark{8,9},
D.~A.~Finley\altaffilmark{10},
R.~J.~Foley\altaffilmark{11,12},
A.~Gal-Yam\altaffilmark{13},
D.~A.~Goldstein\altaffilmark{14,15},
S.~Gonz\'{a}lez-Gait\'{a}n\altaffilmark{16,17},
R.~R.~Gupta\altaffilmark{18},
D.~A.~Howell\altaffilmark{19,20},
C.~Inserra\altaffilmark{21},
R.~Kessler\altaffilmark{22,23},
C.~Lidman\altaffilmark{24},
J.~Marriner\altaffilmark{10},
P.~Nugent\altaffilmark{14,15},
T.~A.~Pritchard\altaffilmark{7},
M.~Sako\altaffilmark{25},
S.~Smartt\altaffilmark{21},
R.~C.~Smith\altaffilmark{26},
H.~Spinka\altaffilmark{18},
R.~C.~Thomas\altaffilmark{15},
R.C.~Wolf\altaffilmark{25},
A.~Zenteno\altaffilmark{26},
T. M. C.~Abbott\altaffilmark{26},
A.~Benoit-L{\'e}vy\altaffilmark{27,28,29},
E.~Bertin\altaffilmark{27,29},
D.~Brooks\altaffilmark{28},
E.~Buckley-Geer\altaffilmark{10},
A.~Carnero~Rosell\altaffilmark{30,31},
M.~Carrasco~Kind\altaffilmark{11,32},
J.~Carretero\altaffilmark{3,33},
M.~Crocce\altaffilmark{3},
C.~E.~Cunha\altaffilmark{34},
L.~N.~da Costa\altaffilmark{30,31},
S.~Desai\altaffilmark{35,36},
H.~T.~Diehl\altaffilmark{10},
P.~Doel\altaffilmark{28},
J.~Estrada\altaffilmark{10},
A.~E.~Evrard\altaffilmark{37,38},
B.~Flaugher\altaffilmark{10},
P.~Fosalba\altaffilmark{3},
J.~Frieman\altaffilmark{10,22},
D.~W.~Gerdes\altaffilmark{38},
D.~Gruen\altaffilmark{34,39,40,41},
R.~A.~Gruendl\altaffilmark{11,32},
D.~J.~James\altaffilmark{26},
K.~Kuehn\altaffilmark{24},
N.~Kuropatkin\altaffilmark{10},
O.~Lahav\altaffilmark{28},
T.~S.~Li\altaffilmark{42},
J.~L.~Marshall\altaffilmark{42},
P.~Martini\altaffilmark{43,44},
C.~J.~Miller\altaffilmark{37,38},
R.~Miquel\altaffilmark{45,33},
B.~Nord\altaffilmark{10},
R.~Ogando\altaffilmark{30,31},
A.~A.~Plazas\altaffilmark{46},
K.~Reil\altaffilmark{40},
A.~K.~Romer\altaffilmark{47},
A.~Roodman\altaffilmark{34,40},
E.~S.~Rykoff\altaffilmark{34,40},
E.~Sanchez\altaffilmark{48},
V.~Scarpine\altaffilmark{10},
M.~Schubnell\altaffilmark{38},
I.~Sevilla-Noarbe\altaffilmark{48,11},
M.~Soares-Santos\altaffilmark{10},
F.~Sobreira\altaffilmark{10,30},
E.~Suchyta\altaffilmark{25},
M.~E.~C.~Swanson\altaffilmark{32},
G.~Tarle\altaffilmark{38},
A.~R.~Walker\altaffilmark{26},
W.~Wester\altaffilmark{10}
\\ \vspace{0.2cm} (The DES Collaboration) \\
}
\email{$\star$ mat.smith@soton.ac.uk}


\begin{abstract}

  We present \taz, a new hydrogen-poor superluminous supernova
  (SLSN-I) discovered by the Dark Energy Survey (DES) supernova
  program, with additional photometric data provided by the Survey
  Using DECam for Superluminous Supernovae (SUDSS). Spectra obtained
  using OSIRIS on the Gran Telescopio CANARIAS (GTC) show \taz\ is a
  SLSN-I at $z=0.608$. Multi-color photometry reveals a double-peaked
  light curve: a blue and relatively bright initial peak that fades
  rapidly prior to the slower rise of the main light curve. Our
  multi-color photometry allows us, for the first time, to show that
  the initial peak cools from 22,000K to 8,000K over 15 rest-frame
  days, and is faster and brighter than any published core-collapse
  supernova, reaching 30\% of the bolometric luminosity of the main
  peak. No physical \nickel-powered model can fit this initial peak.
  We show that a shock-cooling model followed by a magnetar driving
  the second phase of the light curve can adequately explain the
  entire light curve of \taz.  Models involving the shock-cooling of
  extended circumstellar material at a distance of
  $\simeq400$\,\rsolar\ are preferred over the cooling of shock-heated
  surface layers of a stellar envelope. We compare \taz\ to the few
  double-peaked SLSN-I events in the literature. Although the
  rise-times and characteristics of these initial peaks differ, there
  exists the tantalizing possibility that they can be explained by one
  physical interpretation.

\end{abstract}



\section{Introduction}
\label{sec:Intro}

Over the last 10 years, wide-field optical surveys have uncovered a
new class of highly-luminous transients: `superluminous' supernovae
\citep[SLSNe; see review of][]{GalYam2012}.  At $M_\mathrm{peak} <
-21$\,mag, SLSNe are 10-100 times brighter than classical
core-collapse SN events, but are rarer, with $<0.1$\% of the rate
\citep{Quimby2013,McCrum2015,Prajs2016} and only 30 well-studied
examples.  A physical understanding of these extreme events is still
emerging.  There are at least two distinct classes \citep{GalYam2012}:
SLSNe-II show (narrow) hydrogen emission lines, believed to be
generated by interaction with circumstellar material
\citep[CSM;][]{Ofek2007,Smith2007,Gezari2009,Benetti2014}, whereas
SLSNe-I are hydrogen poor \citep{Quimby2011}, and would traditionally
be classified as type Ic SNe.  SLSNe-I almost invariably explode in
galaxies that are low-mass, compact dwarfs
\citep{Neill2011,2015ApJ...804...90L}, and that are metal-poor and
strongly star-forming \citep{Lunnan2013,Chen2013,Leloudas2015}.

The power source of classical SNe Ic, the radioactive decay of
\nickel, cannot easily explain SLSNe-I: several solar masses of
\nickel\ are required to reach $M_\mathrm{peak}<-21$, and the light
curves are difficult to reproduce with a model that has an ejecta mass
greater than the \nickel\ mass \citep[$\mejecta>\mnickel$,
e.g.,][]{Chatzopoulos2009,Chomiuk2011,Inserra2013,Andreas2015}.
Alternative power sources have been proposed, the most popular of
which is energy input from a central engine, such as the spin-down of
a newly-formed and rapidly-rotating magnetar
\citep{Kasen2010,Woosley2010}, or accretion of fallback material onto
a compact object \citep{Dexter2013}.  Interactions between SN ejecta
and a hydrogen-deficient CSM have also been studied
\citep{Chevalier2011,2013ApJ...773...76C}.

A new puzzle has now emerged: some SLSNe-I have double-peaked light
curves, with an early initial peak a few days after the inferred
explosion epoch \citep{2015arXiv151103740N}.  Two events in the
literature have particularly prominent examples: SN2006oz
\citep{2012A&A...541A.129L} and LSQ14bdq \citep{Nicholl2015}.  In both
events, the initial peak lasts a few days in the rest-frame.
\citet{Nicholl2015} modeled the initial peak of LSQ14bdq and found it
inconsistent with being powered by \nickel, favoring instead
luminosity from the shock-heated surface layers of the exploding star
\citep[e.g.][]{2010ApJ...725..904N,Rabinak2011}, followed by reheating
by a central engine to drive the second peak.  In this model, the
brightness and duration of the LSQ14bdq initial peak would imply a
progenitor radius of a few hundred solar radii.  Other explanations,
that can also produce double-peaked light curves, include shock
heating of much more extended material \citep{2015ApJ...808L..51P}, or
a shock breakout driven by a magnetar \citep{2015arXiv150703645K}.

Here, we present \taz, a double-peaked SLSN-I discovered by the Dark
Energy Survey \citep[DES;][]{FlaugherDECam2015} supernova (SN) program
(DES-SN). Lying at a redshift $z=0.608$, the DES-SN $griz$ light
curves probe rest-frame 3000\AA\ to 6000\AA, and provide unique
multi-color information of the initial peak. These multi-color data
allow us, for the first time, to estimate the temperature and
temperature evolution of the initial peak for comparison with various
models. We assume $H_0=70$\,km\,s$^{-1}$\, Mpc$^{-1}$ and a flat
$\Lambda$CDM cosmology with $\Omega_\mathrm{matter}=0.3$ throughout.

\section{Observations}
\label{sec:observations}

\begin{figure*}
\centering
\parbox{0.495\textwidth}{\includegraphics[width=0.495\textwidth]{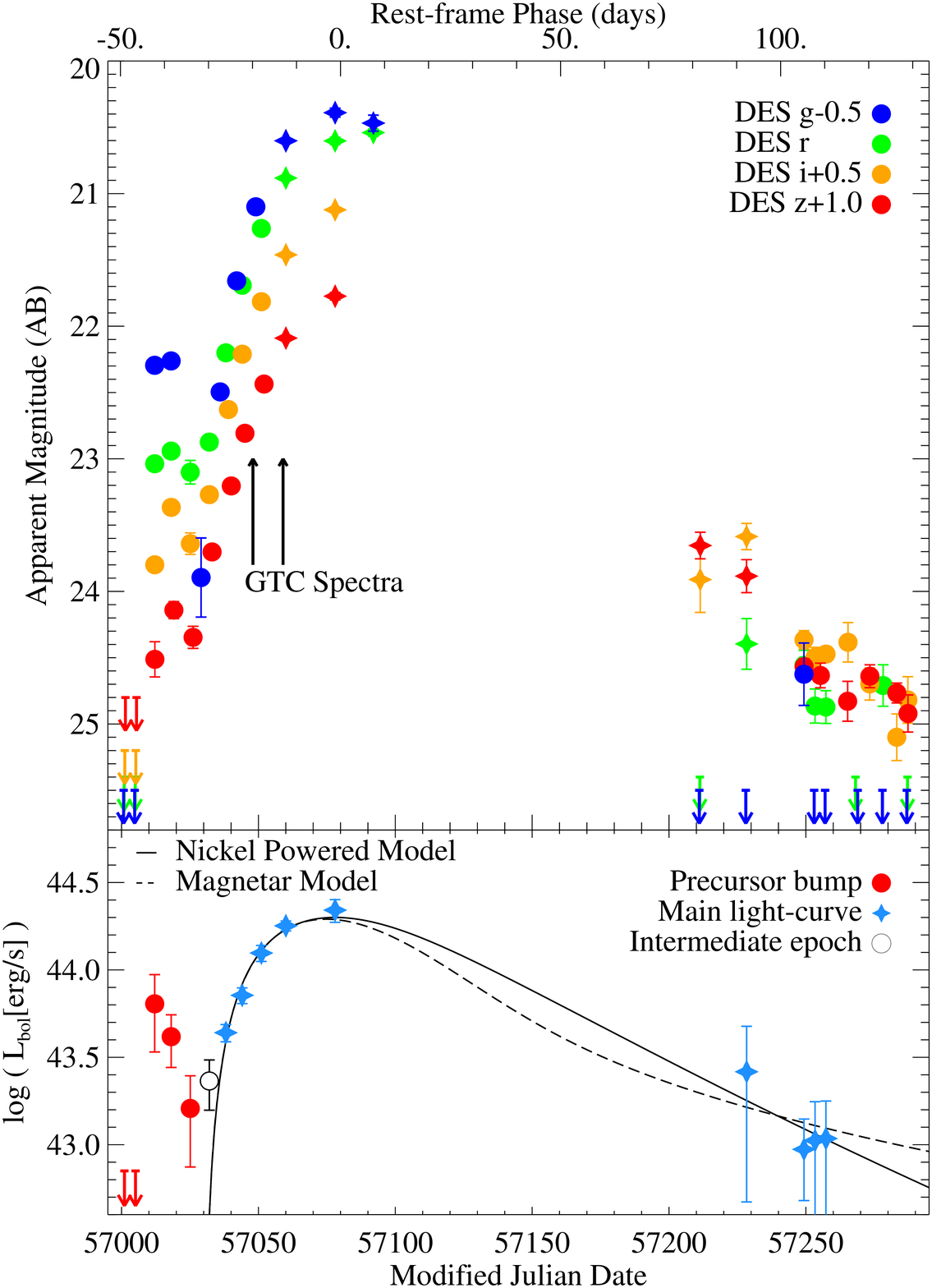}}
\hspace{-0.5cm}\parbox{0.495\textwidth}{\includegraphics[width=0.495\textwidth]{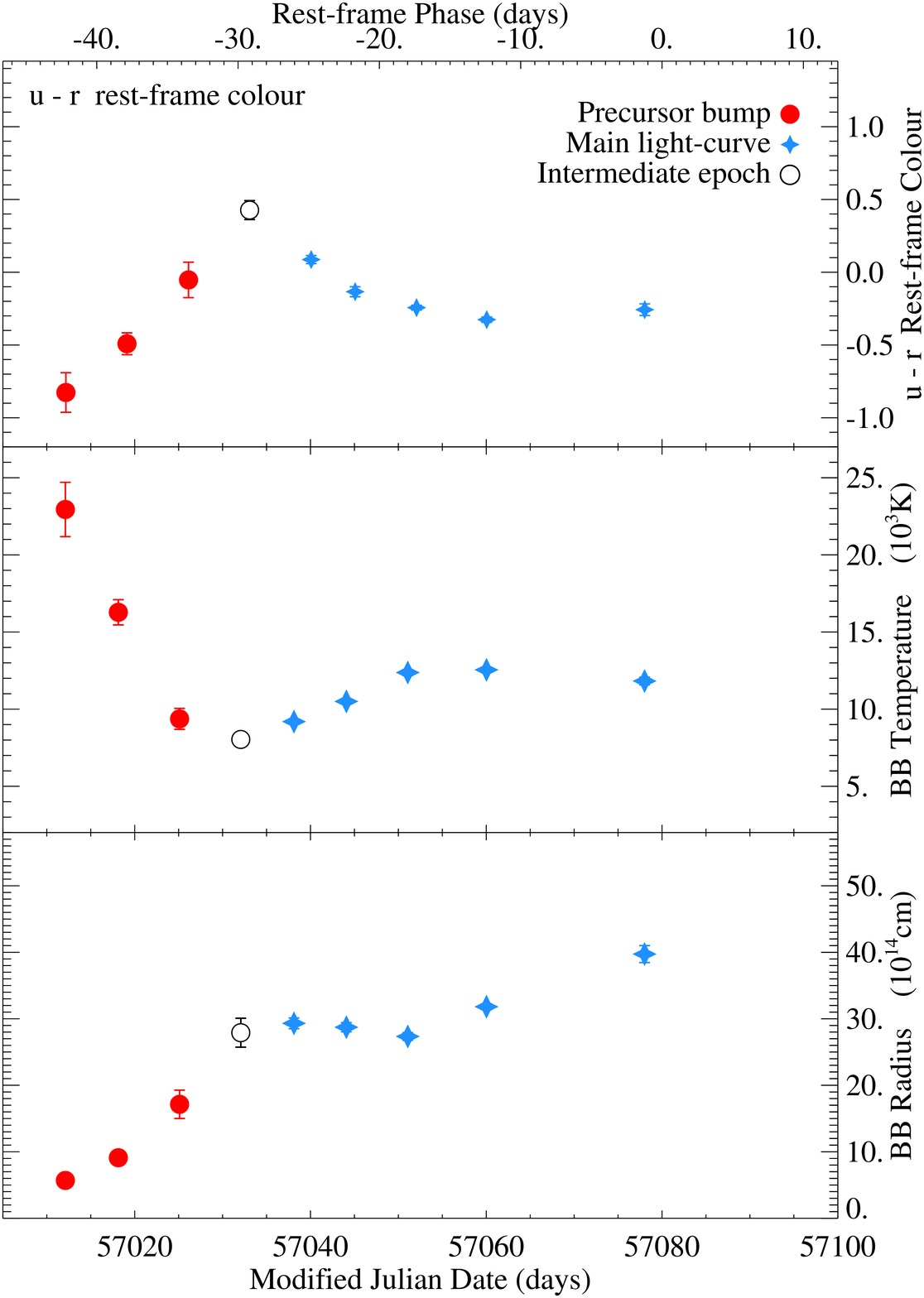}}

\caption{The photometric properties of \taz. Upper Left: The
  multi-color observer-frame $griz$ light curve from DES (circles) and
  SUDSS (crosses). The filters are offset for clarity. Upper limits
  are denoted with arrows, and the epochs of GTC spectroscopy are
  highlighted. Lower left: The bolometric light curve on epochs with 3
  or more filter detections. Epochs associated with the initial peak
  are shown as red circles, with the main light curve plotted as blue
  crosses. Intermediate epochs are plotted as open circles.  The main
  light curve has fits for both a \nickel-powered model (solid;
  $\mnickel=26$\,\msolar) and a magnetar spin-down powered model
  (dashed; $P_{\mathrm{ms}}=2.2$\,ms, $B_{14}=1.3\times{10}^{14}$\,G).
  Upper right: The rest-frame $u$-$r$ color evolution.  Middle and
  lower right: The evolution of the temperature and radius inferred
  from the black-body fits to individual epochs of photometry. Details
  of $k$-corrections and black-body fits are in Section~\ref{sec:analysis}.
\label{fig:x3taz_lc}}
\end{figure*}

\taz\ was discovered as a transient event at $r\simeq23.0$ by DES-SN
in DECam \citep{FlaugherDECam2015} $griz$ images taken on 2014
December 21.1 (all dates UTC) at
$\alpha=02^{\mathrm{h}}28^{\mathrm{m}}04\fs46$,
$\delta=-04\arcdeg05\arcmin12\farcs 7$ (J2000), with previous
non-detections on 2014 December 14.1. The field was imaged again and
\taz\ designated as a transient object on 2014 December 27.1. As the
light curve was monitored in $griz$ by DES over the next 2-3 weeks,
the existence of a pronounced initial peak in the light curve became
clear, and the object was prioritized for spectroscopic observations.
DES monitored \taz\ approximately weekly until February 2015, and
further light-curve data were then obtained via the Search Using DECam
for Superluminous Supernova (SUDSS) project in March 2015 and again in
July 2015. DES re-commenced August 2015, with \taz\ still detected in
the data.  Further details of the DES-SN difference-imaging search
pipeline can be found in \citet{2015AJ....150..172K} and
\citet{2015AJ....150...82G}, with SLSN-specific details in
\citet{Andreas2015}. Photometric measurements were made on the DES-SN
and SUDSS data using the same pipeline as in \citet{Andreas2015},
which has also been used extensively in the literature
\citep[e.g.,][and references therein]{Firth2015}.  The light-curve
data are in Table~\ref{tab:lc_table} and Figure~\ref{fig:x3taz_lc};
they are also available from the WISeREP
archive\footnote{\url{http://wiserep.weizmann.ac.il/}}
\citep{2012PASP..124..668Y}.

\taz\ was spectroscopically observed on 2015 January 26 (when the
target was $r\simeq21.5$) at the Gran Telescopio CANARIAS (GTC) using
the Optical System for Imaging and low-Intermediate-Resolution
Integrated Spectroscopy (OSIRIS) and the R500R grism. The spectrum was
taken in relatively poor conditions (bright sky, 1.1\arcsec\ seeing)
and is of low signal-to-noise (S/N), so was re-observed on 2015
February 6 in dark conditions and 0.8\arcsec\ seeing. The spectra were
reduced using standard Image Reduction and Analysis Facility
(\texttt{IRAF}) routines, and have an effective wavelength coverage of
5200-9000\AA\ (observer frame). The spectra are spectroscopically
similar and available from WISeREP.

\begin{figure*}
\centering{
\hspace{-1.0cm}
\includegraphics[width=1.0\textwidth]{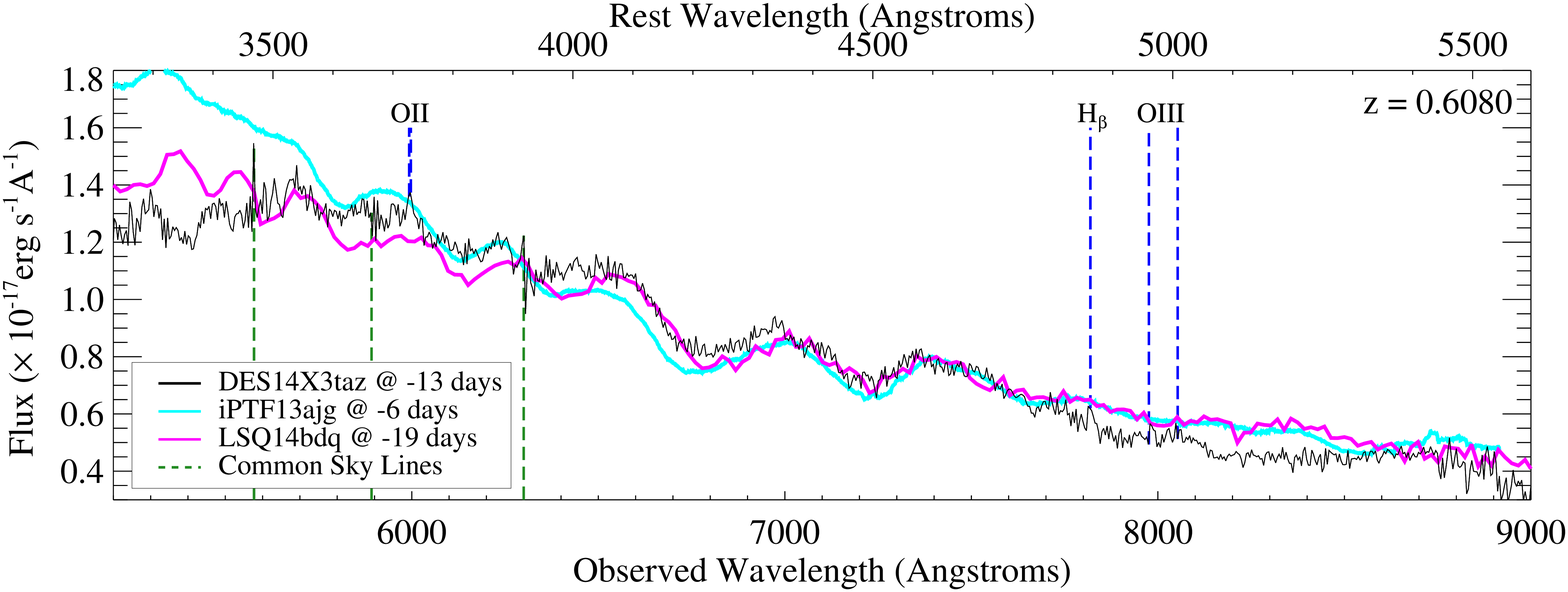}
}
\caption{The GTC optical spectrum of \taz\ from 2015 February 06
  (black line).  Galaxy emission features are highlighted in blue, and
  prominent sky lines in green. Also shown are the spectra of the
  SLSNe-I iPTF13ajg at $-6$~d \citep[cyan;][]{2014ApJ...797...24V},
  and LSQ14bdq at $-19$~d
  \citep[magenta;][]{Nicholl2015}.\label{fig:x3taz_specredshift}}
\end{figure*}

Figure~\ref{fig:x3taz_specredshift} shows the spectrum obtained on
2015 February 6. Weak H$\beta$, [\ion{O}{2}] and [\ion{O}{3}]
host-galaxy emission lines give a redshift of $z=0.608$.  A comparison
with literature spectra using \texttt{SUPERFIT} \citep{Superfit} gives
an excellent match to a hydrogen-poor SLSN.
The broad absorption features at rest-frame 4200\AA\ and 4500\AA\ are
\ion{O}{2}, seen in all SLSNe-I at this phase
\citep[e.g.,][]{Quimby2011,Inserra2013}.

The host galaxy of \taz\ is detected in stacked images from DES that
contain no SN light. Using \texttt{SEXTRACTOR}
\citep{1996A&AS..117..393B}, we measure host-galaxy AB magnitudes
(\texttt{MAG\_AUTO}) of $(g,r,i,z)$=$(26.16\pm0.39$, $25.07\pm0.13$,
$24.95\pm0.13$, $25.00\pm0.18)$ after correcting for MW extinction.
Using the \texttt{Z-PEG} photometric-redshift code
\citep{2002A&A...386..446L}, a \citet{Kroupa2001} initial mass
function, and the redshift fixed at $z=0.608$, we estimate a
host-galaxy stellar mass of $\log(\mathrm{M}/\msolar)=8.0^{+0.4}_{-0.2}$ and a
star-formation rate (SFR) of
$\log(\mathrm{SFR})=-0.8^{+0.6}_{-0.2}$\,\msolar\,yr$^{-1}$.
These properties are consistent with the hosts of other SLSNe
\citep[e.g.][]{Neill2011,Andreas2015,Leloudas2015}.

\section{Analysis}
\label{sec:analysis}

We next study the unusual light curve of \taz. The most striking
feature (Figure~\ref{fig:x3taz_lc}, upper left panel) is the presence
of a first peak in the light curve around MJD 57015 (2014 December
24), prior to the main peak at MJD 57080 (2015 February 27).  The
initial peak is most pronounced in the bluer filters, but is still
distinct in $z$-band.

We estimate the rest-frame $ugr$ light curve from the observer-frame
$riz$ bands, which correspond to a similar wavelength range at
$z=0.608$, using a standard $k$-correction procedure.  With the
limited wavelength coverage of our GTC spectra, pre- and at-max
$k$-corrections are determined using the spectrally-similar event
LSQ14bdq which has coverage down to 3600\,\AA\ \citep{Nicholl2015},
and late-time $k$-corrections using data from PTF12dam at $+171$~d
\citep{Nicholl2013}. A polynomial fit to the rest-frame $g$-band light
curve indicates \taz\ reached a peak brightness of $M_g=-21.39$ on
MJD 57080 after correcting for foreground extinction and assuming zero
internal reddening.  We show the $u-r$ rest-frame color evolution in
Figure~\ref{fig:x3taz_lc} (upper right panel) -- the initial peak is
blue ($u-r\simeq-0.9$) at first, but quickly reddens by nearly
1.5~mag to $u-r\simeq0.5$ (before becoming bluer again during the
main rise to peak brightness). Such reddening behavior is typical of a rapid
decrease in temperature due to cooling.

We analyze this rapid evolution in more detail by fitting a black-body
to each epoch of photometry with observations of at least 3 filters
within 4 days, using Planck's law and a luminosity distance of
3587\,Mpc. (The epoch at MJD 57032 is excluded from these fits, as at
this epoch there is potentially flux from both components of the light
curve.) The evolution of the temperature and radius inferred from
these fits is shown in Figure~\ref{fig:x3taz_lc} (right panels), and
demonstrates a rapid cooling from 22,000K to 8,000K over 15 rest-frame
days. This is followed by a re-heating
phase during the main light curve, which then itself cools as it
approaches the main peak, consistent with other literature on SLSNe-I.

The bolometric light curve (Figure~\ref{fig:x3taz_lc}, lower left
panel) is also constructed by integrating the best-fit black-body
function on each epoch. The two peaks are particularly clear in
bolometric luminosity: one, the initial peak, prior to MJD 57030, and
the other the main peak.

\subsection{The main peak}
\label{sec:main-peak}

We fit the bolometric light curve of the main peak with two models:
energy deposition from the radioactive decay of \nickel, and energy
deposition from the spin-down of a rapidly-rotating magnetar. In both
cases, assuming a spherically symmetric and homologously expanding
ejecta, the bolometric luminosity $L$ as a function of time $t$ since
explosion is \citep{Arnett1982}
\begin{equation}
\label{eq:luminosity_time}
L\left(t\right)=\mathrm{e}^{-(t/\tau_m)^2}\int_0^{t} \frac{2t^\prime}{{\tau_m}^2} P(t^\prime)\mathrm{e}^{(t^\prime/{\tau_m})^2}dt^\prime,
\end{equation}
where $P(t)$ is the power function, the total absorbed power due to
either \nickel\ or a central engine, and $\tau_m$ is the diffusion
time-scale parameter, a function of the ejecta mass (\mejecta),
opacity ($\kappa$) and kinetic energy ($E_k$).  We assume a constant
$\kappa=0.1$\,cm$^{-2}$\,g$^{-1}$ \citep[see discussion
in][]{Inserra2013}.

The power function appropriate for \nickel\ depends on the \nickel\
mass (\mnickel) synthesized in the explosion, $\tau_m$, and the time
of explosion $t_0$ \citep[e.g.,][]{Inserra2013,Andreas2015}. If we
enforce $\mejecta\geq\mnickel$, we obtain a \nickel\ mass of
$\mnickel=26\pm2.6$\,\msolar\ with $\mejecta=\mnickel$ (shown in
Figure~\ref{fig:x3taz_lc}).  Although the fit is statistically
acceptable, with a $\chi^2$ of 4.9 for 6 degrees of freedom (DOF),
such a model with $\mejecta=\mnickel$ is difficult to reconcile with
the observed spectra \citep[see, e.g., the discussion
in][]{Inserra2013}. Enforcing more physical constraints on the ratio
of $\mnickel$ to $\mejecta$ \citep[e.g.][]{Andreas2015} then results
in implied ejecta velocities significantly greater than observed in
the spectra of SLSNe-I: for example, a limit of $\mnickel<0.7\mejecta$
requires velocities of $>$22,000\,km\,s$^{-1}$, compared to
10,000\,km\,s$^{-1}$ measured from the \taz\ spectrum.  Thus, like
other SLSNe-I, the light curve of \taz\ cannot be fit with a
physically plausible \nickel\ diffusion model.

The magnetar power function depends on two parameters: The initial
spin period ($P_\mathrm{ms}$, in milliseconds) and the initial
magnetic field strength ($B_\mathrm{14}$, in $10^{14}$\,G). Assuming
full-trapping of the magnetar radiation, we find
$P_\mathrm{ms}=2.2\pm1.4$\,ms,
$B_\mathrm{14}=1.25\pm0.30\times10^{14}$\,G, and $\tau_m=54.5$~d.
From this we infer an ejecta mass of $\mejecta=9.5$\,\msolar. The
model fit has a $\chi^2$ of 5.4 with 5 DOF. These values are similar
to other SLSNe-I with initial peaks \citep[e.g., LSQ14bdq:
$P_{\mathrm{ms}}=1.7$\,ms and
$B_\mathrm{14}=0.6\times10^{14}$\,G;][]{Nicholl2015} and SLSNe-I in
general \citep[e.g., DES13S2cmm: $P_{\mathrm{ms}}=5.3$\,ms and
$B_\mathrm{14}=1.4\times10^{14}$\,G;][]{Andreas2015}.

\subsection{Modeling the initial peak}
\label{sec:model-init-peak}

We next examine the initial peak in the light curve. We consider both
a \nickel\ power source (for example, from an underlying SN
explosion), and various shock-cooling models.

\subsubsection{\nickel-powered models} 
\label{sec:nick-power-models}

As the main light curve is well-modeled by a magnetar-driven power
source, it is appropriate to consider whether the initial peak in the
light curve is consistent with a normal \nickel-powered SN explosion,
i.e., the SN that gave birth to the magnetar.  We compare the observed
multi-color light curve of the initial peak in \taz\ to various
stripped envelope SNe (SESNe) from \citet{2014ApJS..213...19B} and
\citet{2014AJ....147...99M}. We use the templates of
\citet{karpenka2016}, where 28 events have been modeled using a simple
parametric form allowing them to be placed at arbitrary redshifts
(including $k$-corrections). As SLSNe-I (and \taz) are hydrogen-poor,
we do not consider type II SNe.

None of the SESN templates are consistent with the initial peak in the
\taz\ light curve: the initial peak rises faster, is significantly
bluer at peak, and is 1.5~mag brighter than any of the template
events.  Similarly, fitting the \nickel-powered model from
Section~\ref{sec:main-peak} to the initial peak requires 2.9\,\msolar\
of pure \nickel\ (i.e., \mnickel=\mejecta) and a large inferred energy
of $E_k=4.1\times{10}^{52}$\,erg. Typical core collapse events produce
$\sim0.1$\,\msolar\ of \nickel, and even a complete burning of
1\,\msolar\ of CO to \nickel\ produces only $1\times{10}^{51}$\,erg
\citep{Nicholl2015}. Other power sources are clearly required.

\begin{figure*}[htb]
\centering
\includegraphics[width=0.70\textwidth]{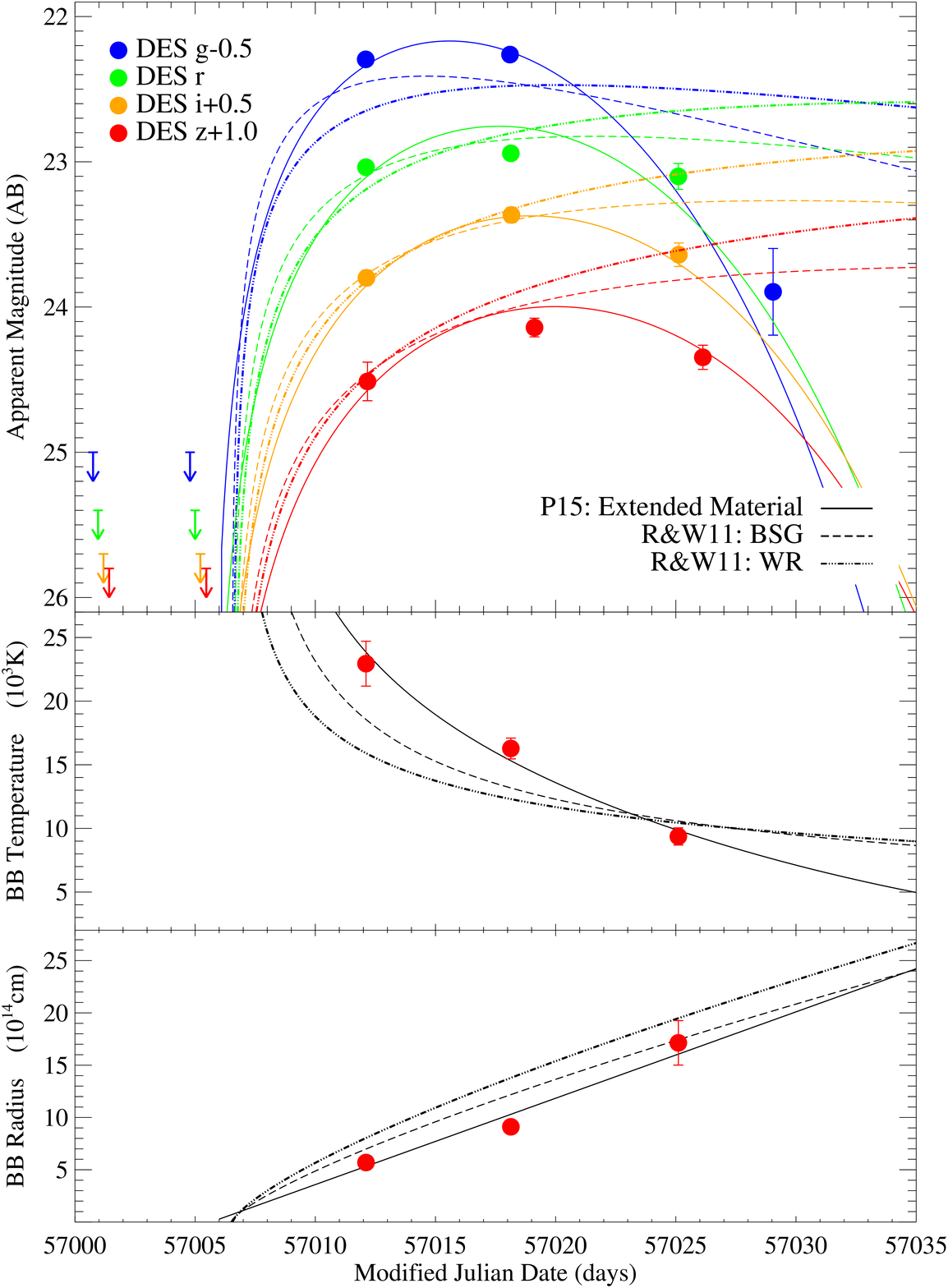}
\caption{ Top: Various analytic models fit to the multi-color data of
  the initial peak of \taz.  The best-fitting shock-cooling models from
  \citet{Rabinak2011} of an extended envelope (BSG; dashed;
  $\mejecta=39$\,\msolar, $R_\star=530$\,\rsolar,
  $E_k=1.5\times{10}^{52}$\,erg) and a compact model (WR; dot-dashed;
  $\mejecta=500$\,\msolar, $R_\star=50$\,\rsolar,
  $E_k=2.5\times{10}^{53}$\,erg), together with an extended material
  model from \citet{2015ApJ...808L..51P} (solid;
  $M_{\mathrm{core}}=9.5$\,\msolar, $M_{\mathrm{ext}}=2.7$\,\msolar,
  $R_\textrm{ext}=400$\,\rsolar,
  $E_\textrm{sn}=6.0\times{10}^{51}$\,erg) are shown. Middle and lower
  panels: The \taz\ temperature and radius evolution from black-body
  fitting (red circles; from Figure~\ref{fig:x3taz_lc}) compared to
  the shock-cooling models discussed above.
  \label{fig:bumpmodelfits}}
\end{figure*}

\subsubsection{Shock cooling of a stellar envelope} 
\label{sec:shock-cool-stell}

We next consider the possibility that the initial peak of \taz\ is
driven by shock cooling where, following a SN shock-wave which heats
the stellar envelope, the envelope expands and cools adiabatically,
releasing energy. We use the analytical models of \citet{Rabinak2011},
and consider various progenitor envelopes: a radiative H envelope
(`blue supergiant', BSG), a convective H envelope (`red supergiant',
RSG), He-dominated envelopes ('He envelopes') and C/O-dominated
envelopes (`Wolf-Rayet').  The \citet{Rabinak2011} models are only
valid at early times, so we cut-off the models according to their
equation 17.  We determine $L(t)$ from their relations, and assuming a
black-body, fit the models to the initial peak of \taz\ by determining
fluxes in the DES band-passes at $z=0.608$.

We show the results in Figure~\ref{fig:bumpmodelfits}. The RSG, BSG
and He envelopes are very similar, so we only show the BSG example for
clarity. Although the rise and color of the initial peak can be
reasonably well matched by these models, in detail the models do not
drop fast enough to match the \taz\ light curve, and do not well
reproduce the inferred temperature evolution, with $\chi^2$'s of 148
and 373 for the BSG and Wolf-Rayet models, respectively (for 8 DOF).
The size of the envelopes implied are also very large (e.g.,
$\simeq520$\,\rsolar\ for the BSG envelope and $\simeq30$\,\rsolar\
for the C/O-dominated envelope). These are a factor of $\sim10$ larger
than typical BSGs or Wolf-Rayet stars.

\subsubsection{Shock cooling of extended material} 
\label{sec:shock-cool-extend}

A similar model involves extended material around the progenitor star,
but at much larger radii than the stellar envelope. If this material
is sufficiently massive and extended -- perhaps from stripping or
earlier pre-explosion eruptions -- the SN shock can propagate into it
and produce a bright initial peak in the light curve
\citep{2010ApJ...724.1396O,2015ApJ...808L..51P} or, in the absence of
further late-time energy input into the ejecta, an isolated bright,
luminous transient \citep{2014ApJ...794...23D}.

We use the analytical relations of \citet{2015ApJ...808L..51P} to
model this, showing the results in Figure~\ref{fig:bumpmodelfits}. Due
to degeneracies in the model between the core mass
($M_{\mathrm{core}}$) and energy, we fix
$M_{\mathrm{core}}=9.5$\,\msolar, the ejecta mass estimated from the
magnetar fitting to the main peak (Section~\ref{sec:main-peak}). The
resultant fit has a $\chi^2$ of 45 for 9 DOF, with
$\simeq$2.7\,\msolar\ of material at a radius of $\simeq400\,\rsolar$
and an energy $E_\textrm{sn}=6.0\times{10}^{51}$\,erg.  (Relaxing this
$M_{\mathrm{core}}$ constraint results in values of
$M_{\mathrm{core}}$ between 5 and 100\,\msolar, but does not
significantly alter the mass of extended material or the $\chi^2$.)
This model reproduces the color, photometric evolution, and
temperature evolution of \taz.

\section{Discussion and Future Work}
\label{sec:discussion}

The discovery of double peaked SLSNe-I provides new insights into the
progenitors of SLSNe. The unique multi-color information of the
initial peak of \taz\ reveals a very hot event that cools rapidly.  We
propose that this is driven by a period of shock cooling, with the
SLSN phase of the light curve driven by re-heating from a central
engine, consistent with a magnetar.  Shock cooling from an extended
stellar envelope can match the color of the early part of the initial
peak, but does not reproduce the fast drop-off, providing a poor
overall match. By contrast, models with extended material at
$\simeq500\rsolar$ can replicate the entire initial peak.

\begin{figure*}
\centering{
\hspace{-0.5cm}\includegraphics[width=1.0\textwidth]{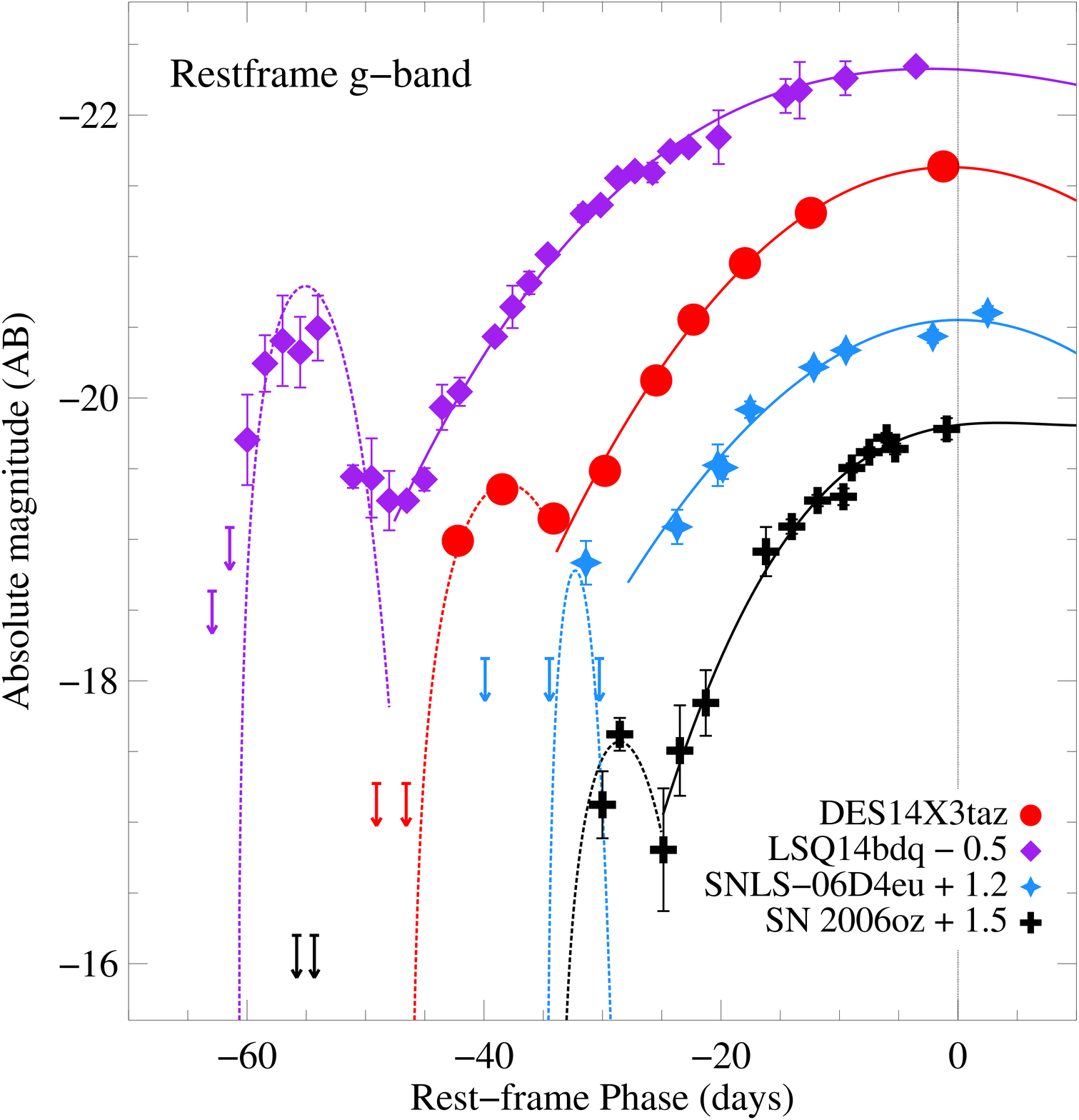}
}
\caption{The rest-frame $g$-band light curve for \taz, compared to literature SLSNe-I with initial peaks. Offsets in absolute magnitude have been applied for clarity. \citet{2015ApJ...808L..51P} model fits are plotted over the initial peaks, and polynomial fits are overplotted on the main peak, of each event. $3\sigma$ non-detections are shown.
  \label{fig:x3taz_restframe}}
\end{figure*}

The quantity of data on double-peaked light curves of SLSNe-I is
increasing quickly. In Figure~\ref{fig:x3taz_restframe}, we compare
the \taz\ rest-frame $g$-band light curve to other SLSNe-I from the
literature that also exhibit initial peaks: SN2006oz
\citep[$z=0.376$;][]{2012A&A...541A.129L}, SNLS-06D4eu
\citep[$z=1.588$;][]{Howell2013} and LSQ14bdq
\citep[$z=0.345$;][]{Nicholl2015}. $griz$ data of the initial peaks
are available for SN2006oz (covering $\simeq3400$\AA--6500\AA\
rest-frame) and SNLS-06D4eu ($\simeq1900$\AA--3400\AA), but only data
from a single broad `$g+r$' filter are available for LSQ14bdq
($\simeq4200$\AA).

Figure~\ref{fig:x3taz_restframe} shows diversity in the luminosity and
duration of the initial peaks, especially when compared to the
rise-time and peak magnitude of the resulting SLSN. However, only
$\simeq15$ literature SLSNe have well-measured pre-explosion
photometry, at least $\simeq30\%$ of which have double peaks. Other
initial peaks may be present in the remainder, but lie below the
photometric detection thresholds of the discovery surveys. The initial
peaks in Figure~\ref{fig:x3taz_restframe} are all consistent with the
\citet{2015ApJ...808L..51P} model, perhaps suggesting that one general
physical interpretation can explain the evolution of all SLSN-I
events.

Understanding the initial peaks of SLSNe-I in more detail will require
spectral (and therefore velocity and composition) measurements. Such
measurements can be used to search for any hydrogen or helium
signatures that may be present, if the model of extended material is
indeed correct.  These observations are technically challenging, as
SLSNe are rare and would need to be identified rapidly from other SN
types based on only 1-2 epochs of photometry.  However, DES is expected to find
$\simeq5-6$ well-observed SLSNe-I per year
\citep{2015arXiv151106670S}, and the distinct properties of these
SLSNe at early times (very blue colors, very faint host galaxies)
allows them to be cleanly selected, making spectral coverage a
realistic possibility in the near future.

\acknowledgments
\begin{center}{\textit{Acknowledgements}}\end{center}

We acknowledge support from EU/FP7-ERC grants 615929 and 307260, STFC,
NSF grant AST-1518052, the Alfred P. Sloan Foundation and a Weizmann-UK 
Grant. Based on observations made with the Gran Telescopio Canarias (GTC), 
at the Spanish Observatorio del Roque de los Muchachos of the Instituto de
Astrofísica de Canarias.

Funding for the DES Projects has been provided by the DOE and NSF(USA), 
MEC/MICINN/MINECO(Spain), STFC(UK), HEFCE(UK). NCSA(UIUC), KICP(U. Chicago), CCAPP(Ohio State), MIFPA(Texas A\&M), CNPQ, FAPERJ, FINEP (Brazil), 
DFG(Germany) and the Collaborating Institutions in the Dark Energy Survey. 
The Collaborating Institutions are Argonne Lab, UC Santa Cruz, University of Cambridge, 
CIEMAT-Madrid, University of Chicago, University College London, DES-Brazil Consortium, 
University of Edinburgh, ETH Z{\"u}rich, Fermilab, University of Illinois, ICE (IEEC-CSIC), 
IFAE Barcelona, Lawrence Berkeley Lab, LMU M{\"u}nchen and the associated 
Excellence Cluster Universe, University of Michigan, NOAO, University of Nottingham, 
Ohio State University, University of Pennsylvania, University of Portsmouth, 
SLAC National Lab, Stanford University, University of Sussex, and Texas A\&M University.

The DES Data Management System is supported by the NSF under Grant
Number AST-1138766. The DES participants from Spanish institutions are
partially supported by MINECO under grants AYA2012-39559,
ESP2013-48274, FPA2013-47986, and Centro de Excelencia Severo Ochoa
SEV-2012-0234. Research leading to these results has received funding
from the ERC including grants 240672, 291329 and 306478.

{\it Facility:} \facility{Blanco, GTC}

\begin{deluxetable*}{lllllll}
\tabletypesize{\footnotesize}
\tablecolumns{7}
\tablewidth{0pt}
\tablecaption{Light curve data for \taz.\label{tab:lc_table}}
\tablehead{
    \colhead{Calendar} &
    \colhead{MJD} &
    \colhead{Phase} &
    \colhead{$f_g$} &
    \colhead{$f_r$} &
    \colhead{$f_i$} &
    \colhead{$f_z$}\\
    \colhead{Date} &
    \colhead{} &
    \colhead{(days)\tablenotemark{a}} &
    \colhead{(counts)\tablenotemark{b}} &
    \colhead{(counts)} &
    \colhead{(counts)} &
    \colhead{(counts)} 
  }
\startdata
10-12-2014 &$57001.1$ & $-49.1$ & $-69.0 \pm 81.1$ & $-56.9 \pm 55.8$ & $0.3 \pm 45.2$ & $-7.9 \pm 67.6$ \\
14-12-2014 &$57005.1$ & $-46.6$ & $-23.3 \pm 68.1$ & $4.3 \pm 45.8$ & $-5.6 \pm 61.7$ & --- \\
14-12-2014 &$57005.2$ & $-46.5$ & --- & --- & --- & $28.6 \pm 90.4$ \\
21-12-2014 &$57012.1$ & $-42.2$ & $1914.4 \pm 61.4$ & $1531.4 \pm 41.8$ & $1203.2 \pm 58.5$ & --- \\
21-12-2014 &$57012.2$ & $-42.2$ & --- & --- & --- & $988.7 \pm 120.9$ \\
27-12-2014 &$57018.1$ & $-38.5$ & $1973.4 \pm 70.1$ & $1672.0 \pm 58.5$ & --- & --- \\
27-12-2014 &$57018.2$ & $-38.5$ & --- & --- & $1792.7 \pm 87.0$ & --- \\
28-12-2014 &$57019.1$ & $-37.9$ & --- & --- & --- & $1391.8 \pm 82.6$ \\
03-01-2015 &$57025.1$ & $-34.1$ & --- & $1444.2 \pm 118.3$ & $1393.4 \pm 104.0$ & --- \\
04-01-2015 &$57026.1$ & $-33.5$ & --- & --- & --- & $1152.0 \pm 88.3$ \\
07-01-2015 &$57029.1$ & $-31.7$ & $438.4 \pm 120.5$ & --- & --- & --- \\
10-01-2015 &$57032.1$ & $-29.8$ & --- & $1779.6 \pm 83.4$ & $1958.4 \pm 69.7$ & --- \\
11-01-2015 &$57033.1$ & $-29.2$ & --- & --- & --- & $2084.6 \pm 77.9$ \\
14-01-2015 &$57036.0$ & $-27.3$ & $1591.4 \pm 59.6$ & --- & --- & --- \\
16-01-2015 &$57038.1$ & $-26.0$ & --- & $3308.9 \pm 44.5$ & --- & --- \\
17-01-2015 &$57039.1$ & $-25.5$ & --- & --- & $3536.7 \pm 63.8$ & --- \\
18-01-2015 &$57040.1$ & $-24.8$ & --- & --- & --- & $3297.5 \pm 70.5$ \\
20-01-2015 &$57042.1$ & $-23.6$ & $3441.5 \pm 71.8$ & --- & --- & --- \\
22-01-2015 &$57044.1$ & $-22.3$ & --- & $5290.8 \pm 113.9$ & $5195.1 \pm 91.5$ & --- \\
23-01-2015 &$57045.1$ & $-21.7$ & --- & --- & --- & $4753.5 \pm 110.3$ \\
27-01-2015 &$57049.1$ & $-19.2$ & $5757.2 \pm 195.3$ & --- & --- & --- \\
29-01-2015 &$57051.1$ & $-18.0$ & --- & $7849.3 \pm 85.4$ & $7481.1 \pm 85.2$ & --- \\
30-01-2015 &$57052.1$ & $-17.4$ & --- & --- & --- & $6690.9 \pm 72.0$ \\
07-02-2015 &$57060.0$ & $-12.4$ & $9105.5 \pm 100.7$ & $11139.1 \pm 97.2$ & $10361.2 \pm 89.6$ & --- \\
07-02-2015 &$57060.1$ & $-12.4$ & --- & --- & --- & $9206.2 \pm 134.1$ \\
25-02-2015 &$57078.0$ & $-1.2$ & $11068.6 \pm 342.2$ & $14419.4 \pm 376.7$ & $14160.5 \pm 263.2$ & $12316.9 \pm 321.3$ \\
11-03-2015 &$57092.0$ & $7.5$ & $10294.3 \pm 574.2$ & $15273.9 \pm 389.2$ & --- & --- \\
08-07-2015 &$57211.4$ & $81.7$ & $775.2 \pm 394.8$ & $552.2 \pm 318.1$ & $1085.7 \pm 247.7$ & $2179.9 \pm 201.8$ \\
25-07-2015 &$57228.3$ & $92.3$ & --- & --- & $1463.7 \pm 133.5$ & $1762.1 \pm 201.5$ \\
25-07-2015 &$57228.4$ & $92.3$ & $165.7 \pm 76.4$ & $438.1 \pm 77.0$ & --- & --- \\
15-08-2015 &$57249.3$ & $105.3$ & $224.0 \pm 48.5$ & $377.7 \pm 39.1$ & $714.3 \pm 45.7$ & --- \\
15-08-2015 &$57249.4$ & $105.3$ & --- & --- & --- & $940.0 \pm 52.7$ \\
19-08-2015 &$57253.4$ & $107.8$ & $66.8 \pm 39.5$ & $284.5 \pm 33.6$ & $633.2 \pm 43.3$ & --- \\
21-08-2015 &$57255.3$ & $109.0$ & --- & --- & --- & $884.0 \pm 75.8$ \\
23-08-2015 &$57257.3$ & $110.3$ & $96.5 \pm 39.4$ & $282.6 \pm 32.2$ & --- & --- \\
23-08-2015 &$57257.4$ & $110.3$ & --- & --- & $646.9 \pm 34.3$ & --- \\
31-08-2015 &$57265.3$ & $115.2$ & --- & --- & --- & $738.9 \pm 102.5$ \\
31-08-2015 &$57265.4$ & $115.3$ & --- & --- & $702.4 \pm 96.1$ & --- \\
03-09-2015 &$57268.2$ & $117.1$ & --- & $160.4 \pm 186.7$ & --- & --- \\
04-09-2015 &$57269.2$ & $117.7$ & $245.0 \pm 188.4$ & --- & --- & --- \\
08-09-2015 &$57273.4$ & $120.2$ & --- & --- & $525.8 \pm 58.8$ & $879.2 \pm 69.8$ \\
13-09-2015 &$57278.2$ & $123.3$ & --- & $328.2 \pm 47.3$ & --- & --- \\
13-09-2015 &$57278.3$ & $123.3$ & $89.6 \pm 50.3$ & --- & --- & --- \\
18-09-2015 &$57283.2$ & $126.4$ & --- & --- & $363.2 \pm 58.9$ & $782.6 \pm 54.2$ \\
22-09-2015 &$57287.2$ & $128.8$ & $452.9 \pm 210.8$ & $178.4 \pm 63.9$ & --- & --- \\
22-09-2015 &$57287.3$ & $128.9$ & --- & --- & $470.6 \pm 76.0$ & $678.8 \pm 87.6$ \\
\enddata
\tablenotetext{a}{Relative to maximum light in the rest frame}
\tablenotetext{b}{Fluxes $f$ in each filter are given in counts. A zeropoint of 31.0 converts counts into AB magnitudes.}
\end{deluxetable*}

\section*{Affiliations}
\small{
\noindent$^{1}$ School of Physics and Astronomy, University of Southampton,  Southampton, SO17 1BJ, UK \\
$^{2}$ Institute of Cosmology \& Gravitation, University of Portsmouth, Portsmouth, PO1 3FX, UK \\
$^{3}$ Institut de Ci\`encies de l'Espai, IEEC-CSIC, Campus UAB, Carrer de Can Magrans, s/n,  08193 Bellaterra, Barcelona, Spain \\
$^{4}$ School of Sciences, European University Cyprus, 6 Diogenes Street, Engomi, 1516 Nicosia, Cyprus \\
$^{5}$ School of Physics, University of Melbourne, Parkville, VIC 3010, Australia \\
$^{6}$ George P. and Cynthia Woods Mitchell Institute for Fundamental Physics \& Astronomy, Texas A. \& M. University, Department of Physics and Astronomy, 4242 TAMU, College Station, TX 77843, USA \\
$^{7}$ Centre for Astrophysics \& Supercomputing, Swinburne University of Technology, Victoria 3122, Australia \\
$^{8}$ School of Mathematics and Physics, University of Queensland, QLD 4072 Australia \\
$^{9}$ ARC Centre of Excellence for All-sky Astrophysics (CAASTRO) \\
$^{10}$ Fermi National Accelerator Laboratory, P. O. Box 500, Batavia, IL 60510, USA \\
$^{11}$ Department of Astronomy, University of Illinois, 1002 W. Green Street, Urbana, IL 61801, USA \\
$^{12}$ Department of Physics, University of Illinois, 1110 W. Green St., Urbana, IL 61801, USA \\
$^{13}$ Department of Particle Physics and Astrophysics, Weizmann Institute of Science, Rehovot 76100, Israel \\
$^{14}$ Astronomy Department, University of California at Berkeley, Berkeley, CA 94720, USA \\
$^{15}$ Lawrence Berkeley National Laboratory, 1 Cyclotron Road, Berkeley, CA 94720, USA \\
$^{16}$ Millennium Institute of Astrophysics, Casilla 36-D, Santiago, Chile \\
$^{17}$ Centro de Modelamiento Matemático, Universidad de Chile, Beaucheff 851, edificio norte,  piso 7, Santiago, Chile \\
$^{18}$ Argonne National Laboratory, 9700 South Cass Avenue, Lemont, IL 60439, USA \\
$^{19}$ Las Cumbres Observatory Global Telescope Network, Goleta, CA 93117, USA \\
$^{20}$ Department of Physics, University of California, Santa Barbara, CA 93106-9530, USA \\
$^{21}$ Astrophysics Research Centre, School of Mathematics and Physics, Queen's University Belfast, Belfast BT7 1NN, UK \\
$^{22}$ Kavli Institute for Cosmological Physics, University of Chicago, Chicago, IL 60637, USA \\
$^{23}$ Department of Physics and Astronomy, 5640 South Ellis Avenue, University of Chicago, Chicago, IL 60637, USA \\
$^{24}$ Australian Astronomical Observatory, North Ryde, NSW 2113, Australia \\
$^{25}$ Department of Physics and Astronomy, University of Pennsylvania, Philadelphia, PA 19104, USA \\
$^{26}$ Cerro Tololo Inter-American Observatory, National Optical Astronomy Observatory, Casilla 603, La Serena, Chile \\
$^{27}$ CNRS, UMR 7095, Institut d'Astrophysique de Paris, F-75014, Paris, France \\
$^{28}$ Department of Physics \& Astronomy, University College London, Gower Street, London, WC1E 6BT, UK \\
$^{29}$ Sorbonne Universit\'es, UPMC Univ Paris 06, UMR 7095, Institut d'Astrophysique de Paris, F-75014, Paris, France \\
$^{30}$ Laborat\'orio Interinstitucional de e-Astronomia - LIneA, Rua Gal. Jos\'e Cristino 77, Rio de Janeiro, RJ - 20921-400, Brazil \\
$^{31}$ Observat\'orio Nacional, Rua Gal. Jos\'e Cristino 77, Rio de Janeiro, RJ - 20921-400, Brazil \\
$^{32}$ National Center for Supercomputing Applications, 1205 West Clark St., Urbana, IL 61801, USA \\
$^{33}$ Institut de F\'{\i}sica d'Altes Energies (IFAE), The Barcelona Institute of Science and Technology, Campus UAB, 08193 Bellaterra (Barcelona) Spain \\
$^{34}$ Kavli Institute for Particle Astrophysics \& Cosmology, P. O. Box 2450, Stanford University, Stanford, CA 94305, USA \\
$^{35}$ Excellence Cluster Universe, Boltzmannstr.\ 2, 85748 Garching, Germany \\
$^{36}$ Faculty of Physics, Ludwig-Maximilians University, Scheinerstr. 1, 81679 Munich, Germany \\
$^{37}$ Department of Astronomy, University of Michigan, Ann Arbor, MI 48109, USA \\
$^{38}$ Department of Physics, University of Michigan, Ann Arbor, MI 48109, USA \\
$^{39}$ Max Planck Institute for Extraterrestrial Physics, Giessenbachstrasse, 85748 Garching, Germany \\
$^{40}$ SLAC National Accelerator Laboratory, Menlo Park, CA 94025, USA \\
$^{41}$ Universit\"ats-Sternwarte, Fakult\"at f\"ur Physik, Ludwig-Maximilians Universit\"at M\"unchen, Scheinerstr. 1, 81679 M\"unchen, Germany \\
$^{42}$ George P. and Cynthia Woods Mitchell Institute for Fundamental Physics and Astronomy, and Department of Physics and Astronomy, Texas A\&M University, College Station, TX 77843,  USA \\
$^{43}$ Center for Cosmology and Astro-Particle Physics, The Ohio State University, Columbus, OH 43210, USA \\
$^{44}$ Department of Astronomy, The Ohio State University, Columbus, OH 43210, USA \\
$^{45}$ Instituci\'o Catalana de Recerca i Estudis Avan\c{c}ats, E-08010 Barcelona, Spain \\
$^{46}$ Jet Propulsion Laboratory, California Institute of Technology, 4800 Oak Grove Dr., Pasadena, CA 91109, USA \\
$^{47}$ Department of Physics and Astronomy, Pevensey Building, University of Sussex, Brighton, BN1 9QH, UK \\
$^{48}$ Centro de Investigaciones Energ\'eticas, Medioambientales y Tecnol\'ogicas (CIEMAT), Madrid, Spain \\
}
\end{document}